\documentclass[twocolumn,prl,showpacs,preprintnumbers,amsmath,amssymb,superscriptaddress]{revtex4}
\usepackage{graphicx}
\usepackage{color}

\begin{document}
\title{Field-induced Tomonaga-Luttinger liquid phase of a two-leg spin-1/2 ladder with strong leg interactions}
\author{Tao Hong}
\email[]{hongt@ornl.gov}
\affiliation{Neutron Scattering Science Division, Oak Ridge National Laboratory, Oak Ridge, Tennessee 37831,
USA}
\author{Y. H. Kim}
\affiliation{Department of Physics, University of Florida, Gainesville, Florida 32611, USA}
\author{C. Hotta}
\affiliation{Department of Physics, Faculty of Science, Kyoto Sangyo University, Kyoto 603-8555, Japan}
\author{Y. Takano}
\affiliation{Department of Physics, University of Florida, Gainesville, Florida 32611, USA}
\author{G. Tremelling}
\affiliation{Carlson School of Chemistry and Department of Physics,
Clark University, Worcester, Massachusetts 01610, USA}
\author{M. M. Turnbull}
\affiliation{Carlson School of Chemistry and Department of
Physics, Clark University, Worcester, Massachusetts 01610, USA}
\author{C. P. Landee}
\affiliation{Carlson School of Chemistry and Department of Physics,
Clark University, Worcester, Massachusetts 01610, USA}
\author{H.-J. Kang}
\affiliation{National Institute of Standards and Technology, Gaithersburg, Maryland 20899, USA}
\author{N. B. Christensen}
\affiliation{Laboratory for Neutron Scattering, ETH Zurich $\&$ Paul Scherrer Institute, CH-5232 Villigen PSI,
Switzerland}
\author{K. Lefmann}
\affiliation{Nanoscience Center, Niels Bohr Institute, University of Copenhagen, DK-2100 Copenhagen, Denmark}
\author{K. P. Schmidt}
\affiliation{Lehrstuhl f$\ddot{u}$r theoretische Physik I, TU Dortmund, Otto-Hahn-Stra\ss e 4, D-44221 Dortmund,
Germany}
\author{G. S. Uhrig}
\affiliation{Lehrstuhl f$\ddot{u}$r theoretische Physik I, TU Dortmund, Otto-Hahn-Stra\ss e 4, D-44221 Dortmund,
Germany}
\author{C. Broholm}
\affiliation{National Institute of Standards and
Technology, Gaithersburg, Maryland 20899, USA}
\affiliation{Department of Physics and Astronomy, The Johns Hopkins University, Baltimore,
Maryland 21218, USA}

\date{\today}

\begin{abstract}
We study the magnetic-field-induced quantum phase transition from a gapped quantum phase that has no magnetic
long-range order into a gapless phase in the spin-1/2 ladder compound bis(2,3-dimethylpyridinium)
tetrabromocuprate (DIMPY). At temperatures below about 1~K, the specific heat in the gapless phase
attains an asymptotic linear temperature dependence, characteristic of a Tomonaga-Luttinger
liquid. Inelastic neutron scattering and the specific heat measurements in both phases are in good agreement
with theoretical calculations, demonstrating that DIMPY is the first model material for an $S=1/2$ two-leg spin
ladder in the strong-leg regime.
\end{abstract}

\pacs{75.10.Jm 75.40.Gb 75.50.Ee}
\vskip2pc

\maketitle

Gapped ground states comprising singlet pairs of spins are the prevalent nonmagnetic quantum disordered states
in a variety of antiferromagnetic Heisenberg models \cite{Thier08:4,Senthil04:303,Gia}. Among those models,
two-leg spin-1/2 ladders with antiferromagnetic rung and leg exchanges, $J_{\rm rung}$ and $J_{\rm leg}$, are
the simplest whose ground states are yet non-trivial. These states give way to a Tomonaga-Luttinger liquid
(TLL)---a critical state with fractional $S$=1/2 spinon excitations---at a magnetic-field-driven quantum
critical point (QCP) \cite{Haldane81:14}.

Although the quantum phase transition at such a QCP has been extensively investigated theoretically
\cite{Sachdev94:50,Chitra97:55,Furus99:60,Wang00:84}, there have been few experimental studies because of the
scarcity of real systems with right energy scales. $\rm (C_5H_{12}N)_2CuCl_4$, which was originally thought to
be a ladder material \cite{Chaboussant}, later turned out to be a frustrated three-dimensional antiferromagnet
\cite{Stone}. In IPA-CuCl$_3$ \cite{Garlea07:98,Hong10:81}, long range magnetic order---also known as a
Bose-Einstein condensation of magnons \cite{Thier08:4,Giama99:59}---due to interladder interactions dominates
the magnetic-field region above the QCP. Thus far, the only detailed report of a TLL in a two-leg spin-1/2
ladder has concerned $\rm (C_5H_{12}N)_2CuBr_4$, a strong-rung material with $J_{\rm leg}/J_{\rm
rung}$\,$\approx$\,0.25 \cite{Ruegg08:101,Thiel09:102}. For deeper understanding of ladders, development of new
materials with a wide range of $J_{\rm leg}/J_{\rm rung}$ will be crucial. Of special interest are materials in
the strong-leg regime, $J_{\rm leg}/J_{\rm rung}$\,$>$\,1, since quantum fluctuations are more prominent in this
regime and as a result the singlets will be less localized, a state reminiscent of the resonating valence bond
liquid \cite{White94:73,Xian95:52}.

In this Letter, we investigate a magnetic-field-induced quantum phase transition in $\rm (C_7H_{10}N)_2CuBr_4$,
DIMPY for short, a new material in which the $\rm CuBr_4^{-2}$ radicals form two-leg spin ladders along the
crystallographic \textbf{\emph{a}} axis \cite{Shapiro07:952}. Our inelastic neutron scattering (INS)
demonstrates that this compound is a spin-gapped quantum magnet with excellent one-dimensionality. Our
specific-heat measurements reveal the presence of a TLL phase above the critical field
$H_c$\,=\,3.0(3)~T, with no long-range order at least down to 150 mK. With the aid of
perturbative continuous unitary transformations (PCUT) and state-of-the-art density-matrix renormalization-group
(DMRG) calculations, we determine the strengths of the rung and leg exchanges from the INS results in the gapped
phase and the specific-heat results in the TLL phase with remarkable consistency, confirming that DIMPY is an
ideal $S$=1/2 spin-ladder system in the \emph{strong-leg} regime.

Single crystals of deuterated DIMPY were grown according to the method described in Ref.~\cite{Shapiro07:952}.
Prompt-gamma neutron activation analysis measurements showed that 67\% of hydrogen sites are occupied by
deuterium. The zero field INS experiment was performed on SPINS at NIST with a single crystal of a 3.5 g mass
and a 0.5$^\circ$ mosaic spread. The measurements were made in the (\textit{h},\textit{k},0) and
(\textit{h},0,\textit{l}) reciprocal-lattice planes with a standard helium cryostat. The high-field INS
experiment was performed on RITA II at SINQ, PSI. The sample consisted of two single crystals with a total mass
of 2 g coaligned within 0.6$^\circ$. The sample was oriented with the (\textit{h},0,\textit{l}) plane horizontal
and was cooled in a 13.5 T vertical-field cryomagnet. The data rate was increased by employing a multi-blade
crystal analyzer and a position sensitive detector \cite{Bahl04:226}. A Be (or BeO) filter was placed after the
sample to remove high-order contamination, selecting a final neutron energy of 5.0 (or 3.7) meV. The specific
heat measurements were made with relaxation calorimetry at the NHMFL, Tallahasse, on a single crystal of an 8.2
mg mass in fields up to 18 T applied parallel to the \emph{\textbf{c}} axis.

\begin{figure}[t]
\includegraphics[width=7.4cm]{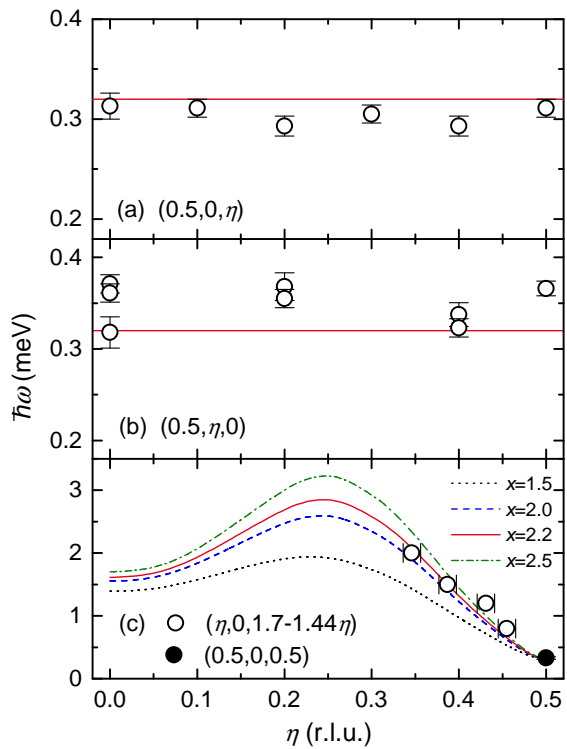}
\caption{(color online). Dispersion measured by INS in DIMPY at $T$\,=\,1.5~K as a function of \textit{h},
\textit{k}, and \textit{l}. Lines in (a) and (b) indicate the gap energy. Lines in (c) are from PCUT
calculations for an AFH two-leg spin ladder for different values of $x$\,=\,$J_{\rm leg}/J_{\rm rung}$.}
\label{dish}
\end{figure}

Figure~\ref{dish} summarizes the zero-field dispersion measured at $T$\,=\,1.5~K by INS along three high
symmetry directions in the reciprocal space \cite{endnote1}. We performed a global fit of all collected data to
a dynamic spin correlation function with the approximate spin-gap dispersion
$\epsilon(\emph{\textbf{q}})$=$\sqrt{\Delta^2+v^2\sin^2[2\pi (0.5-h)]}$ \cite{Masuda06:96}, convolved with
instrumental resolution, finding $\Delta$\,=\,0.32(2)~meV, $v$\,=\,2.36(4)~meV. The individual
data points shown in the figure were obtained by fitting a resolution-corrected lineshape to each
constant-\textbf{\emph{q}} (or constant-energy) scan.  Note that Figs.~\ref{dish}(a) and~\ref{dish}(b) are shown
on a much finer scale than Fig.~\ref{dish}(c). Within a scale as small as 20 $\rm{\mu}$eV, dispersion is absent
along the \emph{\textbf{c}} direction, and only a very weak dispersion, if any, of at most 50 $\rm{\mu}$eV is
found along the \emph{\textbf{b}} direction \cite{endnote2}, indicating that DIMPY is an excellent
one-dimensional (1D) system.

We have calculated the dispersion of an $S$=1/2 antiferromagnetic Heisenberg (AFH) spin-ladder system, using
PCUT \cite{Schmidt05:19} around the limit of isolated rungs. The series in $x$\,=\,$J_{\rm leg}/J_{\rm rung}$ is
obtained in the thermodynamic limit \cite{Knetter03:36} and is extrapolated in terms of an internal parameter
\cite{Schmidt03:34} using Pad\'{e} resummation, yielding reliable results for large $x$ especially for
\textbf{\emph{q}} close to the magnetic zone center. The lines in Fig.~\ref{dish}(c) are the dispersion for
different values of $x$, calculated in conjunction with the accurate gap value $\Delta$\,=\,0.32(2)~meV. Best
agreement with the data is obtained for $x$\,=\,2.2(2), indicating that DIMPY is in the strong-leg regime.

\begin{figure}[t]
\includegraphics[width=8cm]{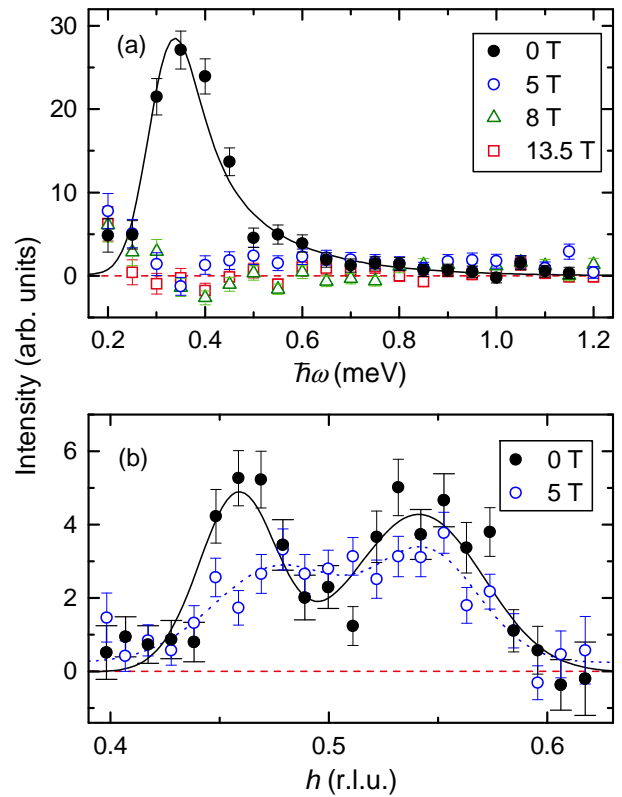}  
\caption{(color online). (a) Background-subtracted constant-\textbf{\emph{q}} scan in DIMPY at the magnetic zone
center (0.5,0,0.9) at $T$\,=\,1.5~K for magnetic fields $H$\,=0,~5,~8, and 13.5 T. (b) Background-subtracted
constant $\hbar\omega$\,=\,0.7 meV scans along the (\textit{h},0,1.7$-$1.44\textit{h})
direction at $T$\,=\,1.5 K and $H$\,=\,0 and 5 T. The dotted line is a guide for the eye. In both frames, solid
lines are fits to a dynamic spin correlation function with the approximate spin-gap dispersion relation
\cite{Masuda06:96}, convolved with instrumental resolution, and dashed lines indicate zero.}
\label{fig:field-neutron}
\end{figure}

Figure~\ref{fig:field-neutron}(a) shows the background-subtracted constant-\textbf{\emph{q}} scan at the
magnetic zone center (0.5,0,0.9) at $T$\,=\,1.5~K in different fields. The background was determined at zero
field by making energy scans at \textbf{\emph{q}}\,=\,(0.35,0,0.9) and (0.65,0,0.9), away from the magnetic zone
center, with the same instrument configuration and by fitting the results to a Gaussian profile over the range
where no magnetic excitation is present. At zero field, the resolution-limited peak indicates the location of
the spin gap. Such a peak is absent at and above 5 T, indicating that the magnetic field drives the system into
a gapless critical phase.

To examine the magnetic excitation spectra at zero field and in the gapless phase, constant-energy scans were
performed at $T$\,=\,1.5 K for $\hbar\omega$\,=\,0.7 meV as shown in Fig.~\ref{fig:field-neutron}(b), where a
constant background term has been subtracted. These measurements were made along the
(\textit{h},0,1.7$-$1.44\textit{h}) direction to maximize the structure factor. The \textbf{\emph{q}}
resolution-limited peaks at zero field are from one-particle excitations. The low-energy feature in the gapless
phase, at 5 T, is clearly much broader than the experimental resolution, suggesting that it arises from a
two-spinon continuum, not from one-particle excitations.

\begin{figure}[t]
\includegraphics[width=8cm]{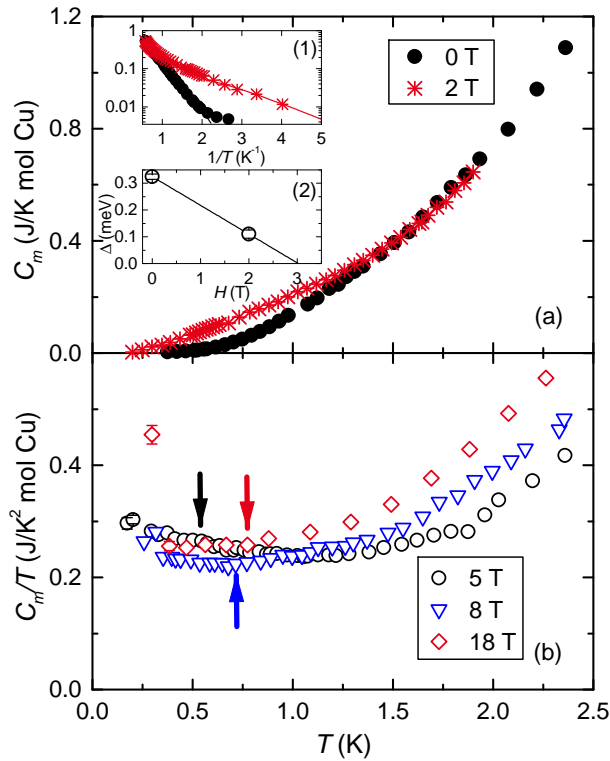} 
\caption{(color online). Magnetic specific heat $C_m$ of DIMPY as a function of temperature $T$ for (a)
$H$\,$<$\,$H_c$ and (b) $H$\,$>$\,$H_c$. In the latter region, the data have been plotted as
$C_m/T$ after subtracting the nuclear-quadrupole contribution (see endnote \cite{endnote3}).
$v_F$ is extracted from data between 0.3~K and the upper limit of the $T$-linear region,
indicated by an arrow. Inset (1): Semilog plot of the $H$\,$<$\,$H_c$ data against $1/T$. Lines are fits
to Eq.~\ref{eq:ct1}. Inset (2): Field dependence of the spin gap obtained from the data.} \label{fig:sc-under-hc}
\end{figure}

To augment the INS results, we measured the specific heat at $T$\,$<$\,2.5~K, as shown in
Fig.~\ref{fig:sc-under-hc}. The phonon contribution was determined from the zero-field entropy $S=\int{(C/T) dT}$ and has been subtracted from the data at all fields. The nuclear-spin contribution has also been subtracted through a simultaneous fit to the data for all fields at temperatures below 700 mK.

At zero field and 2 T, exponentially activated behavior is found, as shown in the first inset to
Fig.~\ref{fig:sc-under-hc}(a), providing additional clear evidence for a spin gap below a critical field. The
specific heat of a gapped 1D AFH quantum magnet in the low-temperature limit is given by \cite{Troyer94:50}:
\begin{eqnarray}
C(T)=\frac{\tilde{n}R}{2\sqrt{2\pi}}\left(\frac{\Delta}{k_B
T}\right)^{3/2}\frac{\Delta}{v}e^{-\Delta/k_B T}, \label{eq:ct1}
\end{eqnarray}
where $\tilde{n}$ is the number of gapped low-energy modes and $R$ the gas constant. Fitted at
$k_BT$\,$\ll$\,$\Delta$ to this expression, the zero-field data yields
$\Delta$\,=\,0.32(1)~meV---excellent agreement with the INS
result---and $\tilde{n}/v$\,=\,1.26(2) \cite{endnote3}. Taking
$v$\,=\,2.36(4) meV from INS, we find $\tilde{n}$\,=\,3.0(1), which unambiguously indicates the
threefold degeneracy expected for a two-leg spin ladder. The field
dependence of $\Delta$ is shown in the second inset to Fig.~\ref{fig:sc-under-hc}; a linear fit
gives $H_{c}$\,=\,3.0(3)\,\,T in good agreement with $\Delta/(g\mu_B)$\,=\,2.8(2)~T,
assuming $g$\,=\,2.0.

Above $H_{c}$, the specific heat shows remarkable behavior. There is no $\lambda$-like peak, indicative of a
phase transition, at temperatures down to 150~mK and magnetic fields up to 18~T. Figure~\ref{fig:sc-under-hc}(b)
shows the magnetic specific heat divided by temperature, $C_m/T$, at 5, 8, and 18~T.
As temperature decreases, $C_m$ reaches an asymptotic
$T$-linear limit, characteristic of TLL, before an upward deviation sets in---probably a
precursor of long-range ordering due to weak interladder interactions \cite{Hagiwara}. The
low-temperature specific heat of TLL is given by conformal field theory as \cite{Blote86:56,Affleck86:56}:
\begin{eqnarray}
C(T)=\frac{\pi}{3}R\frac{k_BT}{v_F(H)}, \label{eq:ct2}
\end{eqnarray}
where $v_F$, the Fermi velocity, is the velocity of the gapless excitations. Using this equation, we extract
$v_F$\,=\,2.79(8), 3.27(11), and 2.89(9)\,meV respectively from the specific heat at 5, 8, and
18~T.

\begin{figure}[t]
\includegraphics[angle=270,width=8cm]{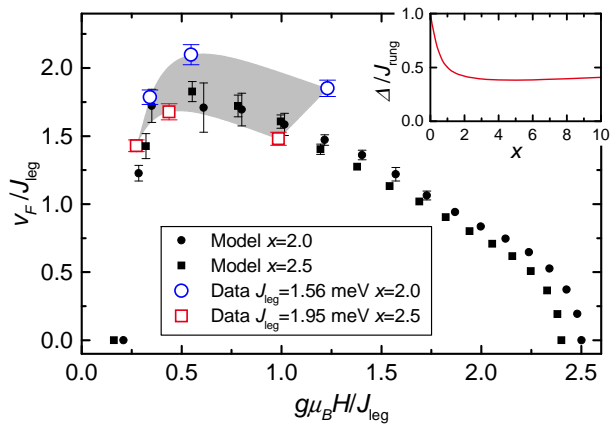} 
\caption{(color online). Field dependence of the Fermi velocity $v_F$ scaled with $J_{\rm
leg}$. Filled symbols are DMRG results for $x$\,=\,2.0 and 2.5. Open symbols are from
specific heat, scaled with $J_{\rm
leg}$\,=\,1.56~meV ($x$\,=\,2.0) and $J_{\rm
leg}$\,=\,1.95~meV ($x$\,=\,2.5), where $J_{\rm leg}$ for each $x$
has been found from calculated $\Delta/J_{\rm rung}$ (see the inset)
and the spin gap $\Delta$\,=\,0.32~meV. For $x$ from 2.0 to 2.5, scaled experimental data
lie in the shaded region.}
\label{fig:sc-above-hc}
\end{figure}

From these $v_F$ and $\Delta$, we now determine $x=J_{\rm leg}/J_{\rm rung}$ and $J_{\rm
rung}$. First, we perform a density-matrix-renormalization-group
(DMRG) calculations for $S$\,=\,1/2 AFH two-leg ladders
\cite{endnote4p5}, in conjunction with finite-size scaling,
and obtain $v_F/J_{\rm leg}$ as a function of $g\mu_{B}H/J_{\rm leg}$ for fixed $x$
\cite{Maeda07:99} and $\Delta/J_{\rm rung}$ as a function of $x$
\cite{endnote5}. From this $\Delta/J_{\rm rung}$ and $\Delta$\,=\,0.32(2)~meV from the
zero-field specific heat and INS, we find $J_{\rm
leg}$---which is $xJ_{\rm rung}$---for
each $x$. With these $J_{\rm leg}$, we then normalize the experimental values of $v_F$ and plot them with the
theoretical results, as shown for $x$\,=\,2 and 2.5 in Fig.~\ref{fig:sc-above-hc}. Finally, comparison of
experiment and theory in this plot yields $x$\,=\,2.3(2), for which $\Delta/J_{\rm rung}$=0.409(6) and thus
$J_{\rm rung}$\,=\,0.78(6)~meV.

To summarize, DIMPY undergoes a quantum phase transition at $H_{c}$\,=\,3.0(3)~T from a gapped
phase to a Tomonaga-Luttinger liquid (TLL). Inelastic neutron scattering reveals the excellent
one-dimensionality of this material and provides a firm value of the spin gap, $\Delta$\,=\,0.32(2)~meV, as does
the specific heat. In the TLL phase, the specific heat attains characteristic $T$-linear
behavior, yielding the Fermi velocity $v_F$ of the gapless excitations for the first time in any
laboratory TLL. We obtain $J_{\rm rung}$\,=\,0.78(6)~meV from $\Delta$\ and $v_F$, and the
exchange ratio $x$\,=\,2.2(2) from the zero-field dispersion and 2.3(2) from
$\Delta$\ and $v_F$. These are consistent with previous estimates, $J_{\rm rung}$\,=\,0.75~meV
and $x$\,=\,1.94, from magnetic susceptibility \cite{Shapiro07:952}. Three independent experiments yielding the
exchange constants with consistency and in excellent agreement with theory establish DIMPY unambiguously as the
first ideal realization of an $S$=1/2 AFH two-leg ladder in the strong-leg regime, thus
opening up an avenue for investigating the properties of such a ladder in this poorly explored regime.

\begin{acknowledgments}
We thank R. Paul for help with neutron activation analysis, and J.-H. Park and T.\,P.~Murphy for help with
cryogenics. The work at ORNL was partially funded by the Division of Scientific User Facilities, Office of BES,
DOE. The work at JHU was supported by the NSF through Grant No. DMR-0306940. CH was supported
by Kakenhi Nos. 19740218, 21110522, and 22014014 from the Ministry of Education, Science, Sports, and Culture of
Japan. KPS acknowledges ESF and EuroHorcs for funding through EURYI. The work at RITA~II, PSI
was supported by the Danish Natural Science Research Council under DANSCATT and by the Swiss NSF contract
PP002-102831. The work at NIST utilized facilities supported in part by the NSF under Agreement No. DMR-0454672.
The cryomagnet at PSI was partially funded by the Carlsberg Foundation. The NHMFL is supported
by NSF Cooperative Agreement No. DMR-0654118, and by the State of Florida and the DOE.
\end{acknowledgments}

\end{document}